\documentclass[journal=ancac3,manuscript=article,layout=twocolumn,]{achemso}
\usepackage[sc]{mathpazo}
\usepackage[scaled=0.9]{helvet}
\usepackage[T1]{fontenc}
\usepackage[utf8]{inputenc}
\usepackage{graphicx}
\usepackage{amsmath,amssymb,amsfonts}
\usepackage{upgreek} 
\usepackage{textcomp}
\usepackage[pdftex,dvipsnames]{xcolor}
\usepackage[pdftex]{hyperref}

\hypersetup{
	pdftitle={Correlated nanoscale analysis of the emission from wurtzite versus zincblende (In,Ga)As/GaAs nanowire core-shell quantum wells},
	colorlinks,
	citecolor=blue
	}

\usepackage{soul}
\usepackage{xspace}

\newcommand{\grad}{$^{\circ}$\xspace}
\newcommand{\celsius}{\,$^{\circ}$C\xspace}

\title{Correlated nanoscale analysis of the emission from wurtzite versus zincblende (In,Ga)As/GaAs nanowire core-shell quantum wells}


\author{Jonas L\"ahnemann}
\affiliation{Paul-Drude-Institut für Festkörperelektronik, Leibniz-Institut im Forschungsverbund Berlin e.V., Hausvogteiplatz 5--7, 10117 Berlin, Germany}
\email{laehnemann@pdi-berlin.de}
\author{Megan O. Hill}
\affiliation{Department of Materials Science and Engineering, Northwestern University, Evanston, Illinois 60208, USA}
\altaffiliation{J. L\"ahnemann and M. O. Hill have contributed equally to this work.}
\author{Jes\'us Herranz}
\affiliation{Paul-Drude-Institut für Festkörperelektronik, Leibniz-Institut im Forschungsverbund Berlin e.V., Hausvogteiplatz 5--7, 10117 Berlin, Germany}
\author{Oliver Marquardt}
\affiliation{Weierstra\ss{}-Institut für Angewandte Analysis und Stochastik, Leibniz-Institut im Forschungsverbund Berlin e.V., Mohrenstr. 39, 10117 Berlin, Germany}
\author{Guanhui Gao}
\affiliation{Paul-Drude-Institut für Festkörperelektronik, Leibniz-Institut im Forschungsverbund Berlin e.V., Hausvogteiplatz 5--7, 10117 Berlin, Germany}
\alsoaffiliation{Present address: Materials Science and Nanoengineering Department, Rice University, Houston, Texas 77005, USA}
\author{Ali Al Hassan}
\affiliation{Naturwissenschaftlich-Technische Fakult\"at der Universit\"at Siegen, 57068 Siegen, Germany}
\author{Arman Davtyan}
\affiliation{Naturwissenschaftlich-Technische Fakult\"at der Universit\"at Siegen, 57068 Siegen, Germany}
\author{Stephan O. Hruszkewycz}
\affiliation{Materials Science Division, Argonne National Laboratory, Argonne, Illinois 60439, USA}
\author{Martin V. Holt}
\affiliation{Center for Nanoscale Materials, Argonne National Laboratory, Argonne, Illinois 60439, USA}
\author{Chunyi Huang}
\affiliation{Department of Materials Science and Engineering, Northwestern University, Evanston, Illinois 60208, USA}
\author{Irene Calvo-Almaz\'an}
\affiliation{Materials Science Division, Argonne National Laboratory, Argonne, Illinois 60439, USA}
\author{Uwe Jahn}
\affiliation{Paul-Drude-Institut für Festkörperelektronik, Leibniz-Institut im Forschungsverbund Berlin e.V., Hausvogteiplatz 5--7, 10117 Berlin, Germany}
\author{Ullrich Pietsch}
\affiliation{Naturwissenschaftlich-Technische Fakult\"at der Universit\"at Siegen, 57068 Siegen, Germany}
\author{Lincoln J. Lauhon}
\affiliation{Department of Materials Science and Engineering, Northwestern University, Evanston, Illinois 60208, USA}
\email{lauhon@northwestern.edu}
\author{Lutz Geelhaar}
\affiliation{Paul-Drude-Institut für Festkörperelektronik, Leibniz-Institut im Forschungsverbund Berlin e.V., Hausvogteiplatz 5--7, 10117 Berlin, Germany}

\begin{document}

\begin{abstract}

While the properties of wurtzite GaAs have been extensively studied during the past decade, little is known about the influence of the crystal polytype on ternary (In,Ga)As quantum well structures. We address this question with a unique combination of correlated, spatially-resolved measurement techniques on core-shell nanowires that contain extended segments of both the zincblende and wurtzite polytypes. Cathodoluminescence hyperspectral imaging reveals a blueshift of the quantum well emission energy by $75\pm15$~meV in the wurtzite polytype segment. Nanoprobe x-ray diffraction and atom probe tomography enable $\mathbf{k}\cdot\mathbf{p}$ calculations for the specific sample geometry to reveal two comparable contributions to this shift. First, there is a 30\% drop in In mole fraction going from the zincblende to the wurtzite segment. Second, the quantum well is under compressive strain, which has a much stronger impact on the hole ground state in the wurtzite than in the zincblende segment. Our results highlight the role of the crystal structure in tuning the emission of (In,Ga)As quantum wells and pave the way to exploit the possibilities of three-dimensional bandgap engineering in core-shell nanowire heterostructures. At the same time, we have demonstrated an advanced characterization toolkit for the investigation of semiconductor nanostructures. 
\end{abstract}

\vspace{10mm}


This study investigates the influence of the crystal phase on the emission characteristics of (In,Ga)As quantum wells (QWs) on GaAs nanowires (NWs). Ternary group-III-arsenide core-shell QWs have shown great promise in NW-based emitters/lasers \cite{Fontcuberta_2008,Dimakis_2014,Koblmuller_2017,Stettner_2018} and detectors/solar cells.\cite{Dai_2014,Treu_2015,Erhard_2015,Moratis_2016} Notably, (In,Ga)As-based emitters have the potential to operate at the so-called telecommunication band;\cite{Kim_2017a} such a combination of nanoscale emitters with Si waveguides promises to revolutionize the speed and energy efficiency of on-chip information transfer.\cite{Chen_2014} Recently, key steps towards nanophotonic on-chip integration of such devices on Si were demonstrated.\cite{Giuntoni_2016, Stettner_2017,Kim_2017} In this context, GaAs NWs act as a substrate for QWs, with the entire structure acting as a waveguide to confine light. 

When GaAs is grown in NW form, one can access both the equilibrium zincblende (ZB) and the metastable wurtzite (WZ) crystal phases,\cite{Koguchi_1992,Persson_2004,Spirkoska_2009,Jahn_2012} and advances in understanding NW growth \cite{Glas_2007,Jacobsson_2016} have recently enabled control over the polytype by both metal-organic vapor phase epitaxy \cite{Lehmann_2012,Jacobsson_2016} and molecular beam epitaxy (MBE).\cite{Rieger_2013} The distinct optical and structural properties of the WZ polytype provide an intriguing opportunity for band gap engineering using polytype junctions.\cite{Caroff_2011} Indeed, studies of binary GaAs NWs have established that the band gap energies of both polytypes are the same to within 5~meV,\cite{Senichev_2018,Vainorius_2018} whereas the WZ/ZB interface exhibits a type II band offset of about 100~meV.\cite{Vainorius_2014,Gurwitz_2012,Spirkoska_2009,Kusch_2014} For binary InAs NWs, a difference between the ZB and WZ band gap energies of about 60~meV has been reported.\cite{Rota_2016} However, little is known on the influence of the crystal phase on the band structure of important ternary alloys such as (In,Ga)As.\cite{Morkotter_2013} Beyond the influence of crystal phase, engineering the optical properties of NW-based QWs requires a detailed understanding of how composition, QW width, and the associated interface strain impact the band structure,\cite{Dimakis_2014, Davtyan_2017,Lewis_2017,Koblmuller_2017,Stettner_2018} necessitating an integrated approach to nanoscale characterization. 

Here we employ a unique combination of spatially-resolved measurement techniques, together with calculations in the framework of $\mathbf{k}\cdot\mathbf{p}$ perturbation theory, to provide a correlative structure-property analysis on the level of single NWs. We grow GaAs NW cores with a WZ segment on a ZB base to overgrow circumferential WZ and ZB (In,Ga)As QWs under identical conditions, exploiting the transfer of the core polytype to the shell.\cite{Paladugu_2009,Popovitz_2011,Corfdir_2016} Using cathodoluminescence (CL) spectroscopy and electron backscatter diffraction (EBSD), we observe a $75\pm15$~meV blueshift of the QW emission energy in the WZ segment. A combination of atom probe tomography (APT), coherent x-ray nanodiffraction, and theory enables two dominant contributions to the blue-shift to be identified. First, the In content of the WZ QW is lower than that of the ZB QW, while the QW thickness remains unchanged. Second, compressive strain in the QW induces distinct shifts in the valence band energy in the WZ and ZB segments. Our findings can guide future work on NW core-shell heterostructures for a range of applications that exploit quantum confinement and polytype phase engineering.

\begin{figure*}[t]
\includegraphics*[width=17.7cm]{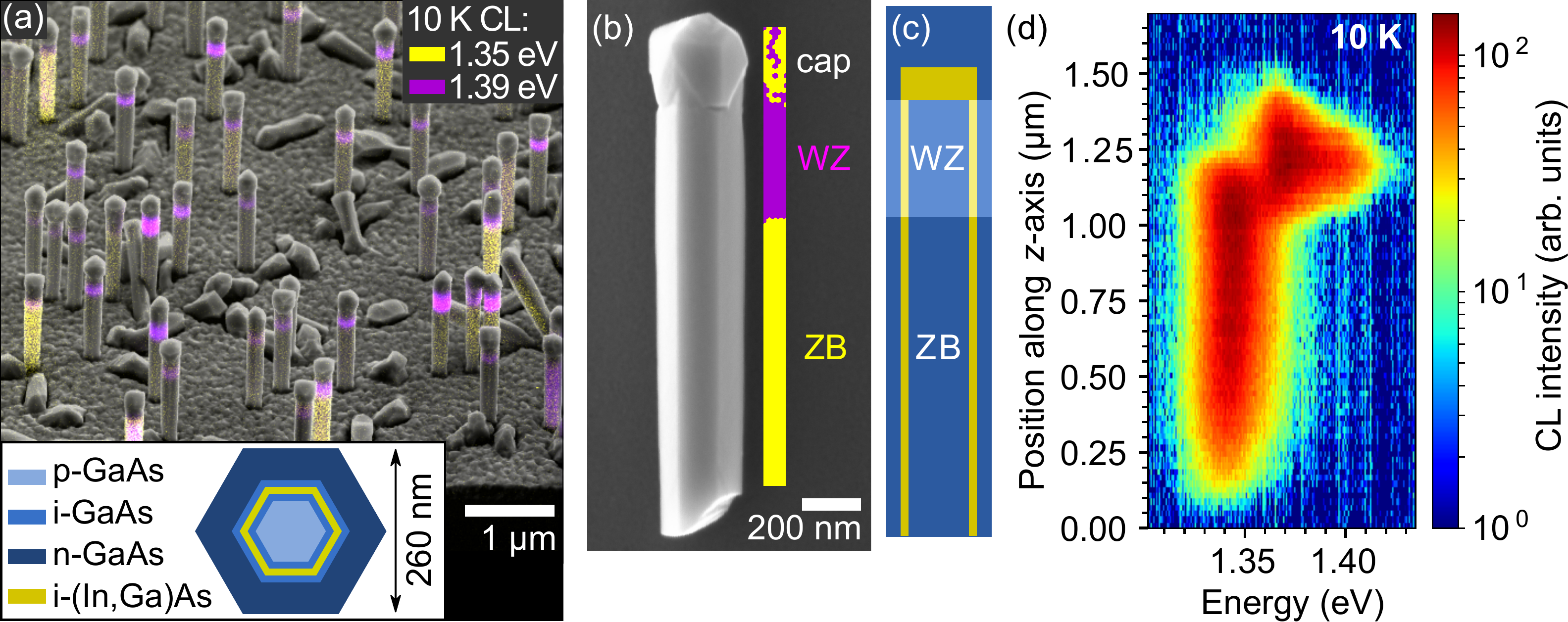}
\caption{\label{fig:intro} (a) Birds-eye view scanning electron micrograph of the (In,Ga)As core-shell NW ensemble (sample~A). Superimposed are monochromatic CL maps of the low-temperature QW emission at 1.35 and 1.39~eV (false color representation using linear intensity scales). The inset shows a top-view illustration of the core-shell geometry. (b) Micrograph of NW A1 together with a map of the crystal phase recorded by EBSD on the upper facet of this NW. (c) Sketch of the axial NW cross-section highlighting how the shell QW is grown on the core segments with different crystal structure. (d) CL spectral line-scan along the NW axis for the NW shown in (b). The emission intensity is color-coded on a logarithmic scale.}
\end{figure*}

\section{Results and Discussion}

GaAs NWs were grown by MBE using the Ga-assisted vapor-liquid-solid growth mode,\cite{Colombo_2008, Jabeen_2008} and then radially overgrown with shells to form heterostructures.\cite{Dimakis_2014} As illustrated in the inset to Fig.~\ref{fig:intro}(a) for sample A, a p-doped GaAs core is followed by an (In,Ga)As QW ($\approx15$\% In) sandwiched between two nominally undoped (intrinsic) GaAs layers. The QW is capped with an n-doped GaAs outer shell, resulting in a facet-to-facet diameter of about $260$~nm. For measurements that require smaller diameter NWs, sample B was grown with a reduced core and capping layer thickness (without doping), leading to an overall NW diameter of about $145$~nm. An overview of the as-grown NW ensemble A is given by the scanning electron micrograph in Fig~\ref{fig:intro}(a); superimposed are two monochromatic CL maps of the QW emission for energies of 1.35 and 1.39~eV, which highlight the presence of segments emitting at different energies (see also discussion and complementary panchromatic CL map in the \textcolor{magenta}{Supporting Information}). The origin of this shift in emission energy and its correlation with the crystal structure are investigated in the current study.

\textbf{Correlation of crystal structure and emission properties}. The crystal structure of the NW heterostructure was investigated with spatially-resolved EBSD and x-ray nano-diffraction, correlated directly with CL, and independently confirmed by transmission electron microscopy (TEM). Firstly, the variations in crystal structure along the NW were determined by EBSD in a SEM for single, dispersed NWs of sample A, mapping the crystal structure and orientation across the central part of the upward-facing side facet. While the crystal orientation is constant, we consistently observe a transition from ZB to WZ polytype as exhibited by the representative NW in Fig.~\ref{fig:intro}(b). The base of the NW consists of an about $1~\upmu$m long ZB segment, followed by about 300--400~nm of the WZ polytype. The cap of the NW ($\approx300$~nm) exhibits a more complex faceting in the micrograph and an indeterminate structure due to an insufficient quality of the Kikuchi patterns in the EBSD measurements. Figure~\ref{fig:intro}(c) schematically illustrates the structure, where the QW shell (yellow) and capping layer are known to adopt the crystal structure of the NW core.\cite{Paladugu_2009,Popovitz_2011,Corfdir_2016}

The ZB to WZ transition occurs due to a change in growth conditions. The ZB polytype is grown with a relatively low V/III flux ratio.\cite{Cirlin_2010, Spirkoska_2009} Towards the end of the core growth, the Ga flux is terminated, leading to the consumption of the Ga droplet. \cite{Yu_2012,Dastjerdi_2016} As the droplet shrinks, the contact angle to the NW sidewalls is reduced, leading to the nucleation of the WZ polytype.\cite{Jacobsson_2016, Plissard_2010} Finally, the cap region results from axial growth during the deposition of the radial shells. The cap is non-emissive and therefore not the focus of the present study. 
 
The crystal structure was spatially correlated with the (In,Ga)As QW light emission from the same NW using low-temperature CL spectroscopy. Figure~\ref{fig:intro}(d) presents a spectral line-scan along the axis of the NW in Fig.~\ref{fig:intro}(b). Emission bands are observed at 1.345 and 1.375~eV, and their intensity maxima coincide with the locations of the ZB and WZ segments, respectively. Therefore, the two bands originate from QWs with different crystal structures. 
A much weaker CL signal around 1.48~eV (not shown), corresponding to the GaAs core, is observed along both the ZB and WZ segments, which confirms the efficient carrier transfer from the core to the QW independent of the crystal polytype. The upper end of the NW does not show any significant luminescence, indicating strong non-radiative recombination in this part of the NW. 

\begin{figure*}[t!]
\centerline{\includegraphics*[width=17.7cm]{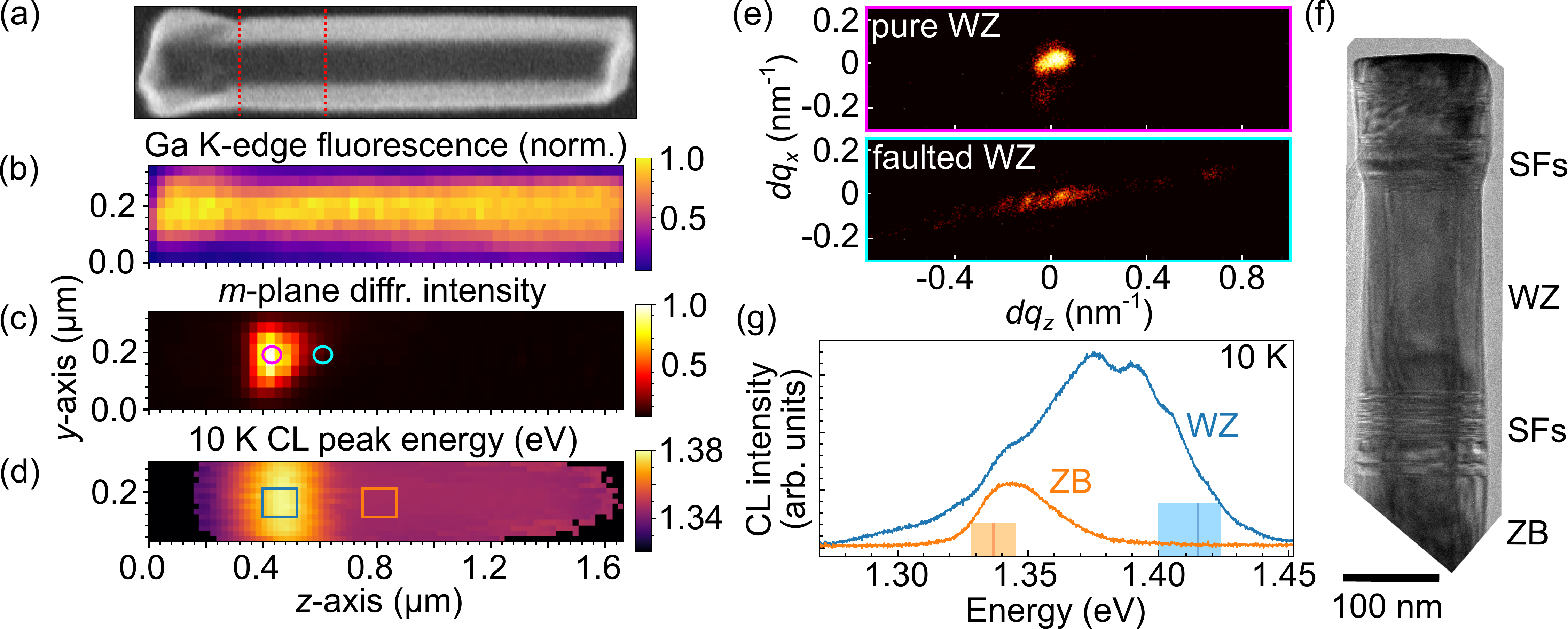}}
\caption{\label{fig:xrd1}(a) SEM micrograph and (b) XRF map of the Ga K-edge for NW A2 used to align the CL and x-ray maps. (c) 2D map of the normalized integrated diffraction intensity for the $(10\bar{1}0)$ reflection (\emph{m}-plane) forbidden in ZB. (d) Map of the CL peak energy for the same NW extracted from a Gaussian fit to the individual spectra of a 2D spectral image. (e) Diffraction patterns for the two positions marked in (c) representing the pure WZ (magenta) and the faulted region (cyan) at the WZ/ZB transition plotted using relative scales. (f) Representative TEM micrograph for a NW from sample B (NW B1) optimized for contrast from stacking defects. The presence of the QW is evidenced by its strain contrast. (g) Exemplary CL spectra for the QW emission on the WZ and ZB segments integrated over the regions marked in (d). The markers highlight our estimate of the energy difference between the QW emission from pure WZ and ZB segments.}
\end{figure*}

Before discussing the CL results in more detail, we consider a more in-depth analysis of the crystal structure that we obtained from synchrotron-based x-ray nano-diffraction (nanoXRD) measurements. Figures~\ref{fig:xrd1}(a) and (b) show a SEM micrograph and a map of the Ga-K edge x-ray fluorescence (XRF) (taken in conjunction with nanoXRD) for the same representative NW from sample A, enabling spatial alignment of the CL and XRD measurements. The integrated diffraction intensity of one of the measured Bragg conditions, $(10\bar{1}0)$, is presented in Fig.~\ref{fig:xrd1}(c). The equivalent to the WZ $(10\bar{1}0)$ reflection is forbidden for the ZB polytype, allowing us to map the position of the WZ segment (about 200--300 nm long) in the NW. Figure~\ref{fig:xrd1}(d) shows a map of the CL peak energy, where again a blueshift of the peak emission from 1.345~eV in ZB to 1.380~eV in the WZ segment is obtained.

In addition, nanoXRD is sensitive to the presence of defects in the crystal stacking order of the WZ region. Fig.~\ref{fig:xrd1}(e) shows distinctive diffraction patterns originating from the center of the WZ segment and close to the ZB interface, as marked in Fig.~\ref{fig:xrd1}(c). The diffraction pattern for the center of the WZ region (magenta) appears as a sharp isolated peak indicative of a pure, or nearly pure, WZ crystal. In the transition region close to the ZB interface (cyan), the diffraction spreads to high $q_{y}$ due to the interference between stacking variations that are smaller than the size of the projected beam.\cite{Hill_2018, Davtyan_2016, Favre-nicolin_2010} This structure was observed in all eight NWs measured by nanoXRD. Indeed, a high density of short, alternating segments of WZ and ZB stacking, including stacking faults (SFs) and twins, have been reported for the transition from the ZB to the WZ phase.\cite{Jabeen_2008,Plissard_2010,Cirlin_2010} At the same time, the central part of the WZ segment appears to be mostly free from SFs, as independently confirmed in Fig.~\ref{fig:xrd1}(f) by TEM for a NW of sample B.

Figure~\ref{fig:xrd1}(g) displays CL spectra for the WZ and ZB segment, highlighting that the QW emission from the WZ segment is blueshifted, broadened and can be more intense with respect to the ZB region. From Fig.~\ref{fig:intro}(d), we further see that the broadening and blueshift increase from the interface between WZ and ZB to the center of the WZ segment. For the interpretation of the CL measurements, it is important to keep in mind that carriers are excited locally within the scattering volume of the electron beam, but the light is collected from a much larger area. Thus, carriers diffusing to and recombining at a region with lower emission energy will still be attributed to the beam position. Normally, even for QW structures, the III-arsenides have a sufficiently large diffusivity that most carriers excited in the WZ segment should reach the lower energy ZB QW.\cite{Hillmer_1989, Pieczarka_2017} Nevertheless, we see a clearly blueshifted transition for the WZ QW. Apparently, the stacking defects in the transition region serve as a barrier limiting carrier diffusion to the lower energy segment. 

We note that the centroids of the WZ and ZB CL bands are separated by 40~meV, though the sources of broadening and line shapes of these peaks merit a careful consideration (discussed here), and motivate verification via experiment-informed $\mathbf{k}\cdot\mathbf{p}$ calculations (detailed in the next sections). In this material system, the QW emission is inhomogeneously broadened from fluctuations in the composition of the ternary alloy and of the QW width, which introduce carrier localization centers in the QW plane. A significant additional broadening is also present in our data, evidenced by the high energy tail of the ZB band and from the multiplicity of peaks when exciting the WZ segment. The different origins of the broadening in the two polytypes must be considered to infer the blueshift one would expect from pure segments.

Firstly, the peak of the ZB band is shifted towards higher energies due to the high excitation density of the electron beam in CL, which results in the saturation of localized states. The additional contributions probably even include higher excited states in the QW. Therefore, we attribute the low energy slope of this broadened emission band to the QW emission in the ZB segment, as highlighted by the orange marker in Fig.~\ref{fig:xrd1}(g).

Secondly, the strong inhomogeneous broadening in the WZ emission band can be attributed to carriers diffusing to the defective region between the two segments. It has been observed that SFs constitute crystal phase QWs,\cite{Spirkoska_2009,Heiss_2011,Jahn_2012} and when superimposed with a compositional core-shell QW, can act as quantum rings, resulting in an up to 50 meV red-shift of the emission with respect to the compositional QW.\cite{Corfdir_2016} Carrier localization at the quantum rings could also explain the increase in emission intensity in the WZ as compared to the ZB segment. The significant broadening of the WZ band is therefore consistent with the red-shift from carriers diffusing to the quantum rings, indicating that the emission of the QW in the pure WZ region produces the higher energy slope, as highlighted by the blue marker in Fig.~\ref{fig:xrd1}(g). Note that, in contrast to ZB, the carrier diffusion to the quantum rings effectively reduces the excitation density in the WZ QW.

Taking these effects into account, we estimate that the WZ QW emission is blueshifted by $75\pm15$~meV with respect to that of the ZB QW. This estimate is based on inferring the energies of the QW emission lines for WZ and ZB that are unbroadened and unshifted in a manner consistent with previously observed phenomena. Importantly, the validity of this analysis can be tested and deeper insights can be attained by invoking further experiments, analyses, and calculations, as presented in the following. 

\begin{figure*}[t!]
\centerline{\includegraphics*[width=18cm]{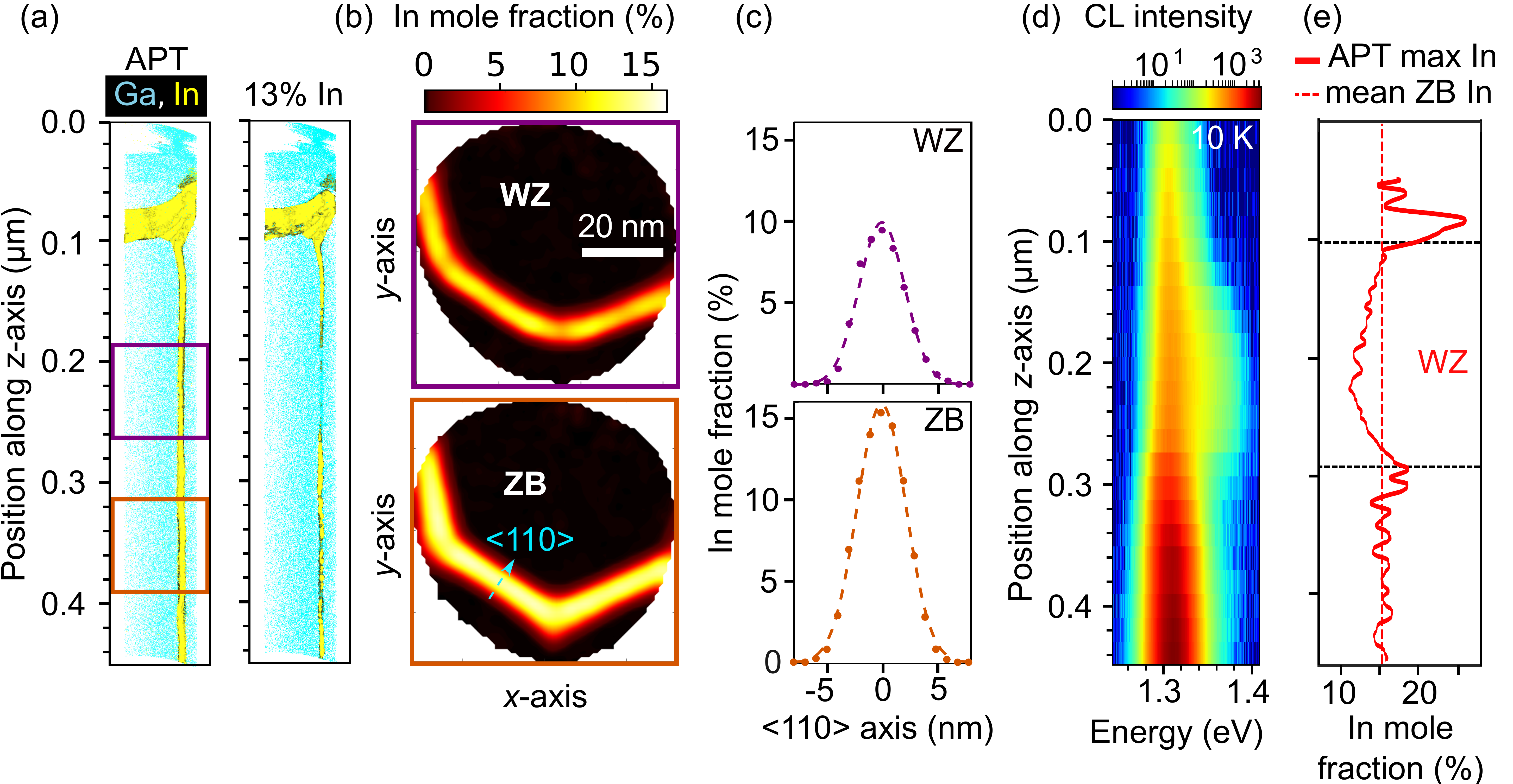}}
\caption{\label{fig:apt}(a) APT data for one side facet of a NW from sample B (NW B2) represented as two-dimensional cut along the NW axis through the APT data on the NW side facet with color-coded composition. The map shows 0.5\% of Ga atoms in blue, and 100\% of In atoms in yellow, while As atoms are omitted for clarity. The outline of the QW is highlighted by the isosurface for a mole fraction of 8\%. In the right panel, the isosurface for a mole fraction of 13\% In is mapped and filled in yellow. (b) Top-view cross sections of the In mole fraction extracted from the APT data for the WZ and ZB segments as marked in (a). (c) 2D slices in the center of the QW in the $\langle110\rangle$ direction, indicated by the dashed line in (b). Representative line profiles are shown for both WZ and ZB, with Gaussian fits to extract the maximum (dashed curve). (d) CL spectral line-scan recorded along the axis of NW B2. (d) Plot of the maximum In mole fraction, determined along the length of the NW by averaging maxima of the QW profiles exampled in (c). This profile was used to determine the In contents used in emission simulations.}

\end{figure*}

\textbf{Influence of the polytype on the QW growth.} The shift in emission energy between the QWs on the WZ and ZB NW segments could arise from differences in the band structure of the two polytypes. \cite{Bechstedt_2013,Marquardt_2017} However, the difference in crystal structure could modify the growth of the QW, resulting in a difference in QW composition, thickness, or strain state, each of which would influence the emission. Here, either thinner QWs or a lower In content on the WZ segments would lead to a blueshift as observed. In order to deconvolve the influence of QW composition and morphology from that of the crystal structure, APT measurements, which provide a three-dimensional, spatially-resolved view of the NW heterostructure, were correlated directly with CL. 

Sample B, which has a reduced diameter but the same evolution of crystal structure [\emph{cf.} Fig.~\ref{fig:xrd1}(f)], was used for APT, as the large diameter of the NWs from sample A would require very high voltages for evaporation leading to a high probability of fracture. As APT is destructive, CL measurements were carried out prior to the APT analysis. NWs were isolated using a tungsten micro-manipulator tip in a dual-beam SEM. In order to align the limited radial field of view of APT with the QW region, the NW was intentionally tilted when mounting (see \textcolor{magenta}{Supporting Information}), such that mainly one side facet of the NW is probed. This specific geometry allowed us to probe the QW along a length of 400--500~nm and thus measure both the WZ and ZB regions in a single wire. A 2D slice in the center of the reconstructed NW is shown in Fig.~\ref{fig:apt}(a) (for a 3D animation of the APT data, see \textcolor{magenta}{Supporting Information}). At the top of the NW, axial GaAs and (In,Ga)As segments are visible, corresponding to the lengthening of the NW during growth of the QW and outer shell layers. Below this region, a single facet of the (In,Ga)As QW can be seen. To further outline this QW, the isosurfaces corresponding to a mole fraction of 8\% In are highlighted. The right panel in Fig.~\ref{fig:apt}(a) shows an isosurface contour for 13\% In mole fraction, in which the interior of the contour has been filled in (yellow). 
The gap in the isosurface indicates that there are compositional variations along the length of the QW. Specifically, the In mole fraction is reduced in the upper part of the probed QW. The purple box coincides with the region of higher emission energy in CL, as shown in Fig.~\ref{fig:apt}(d), suggesting it corresponds to the WZ segment of the NW.

Cross-sectional views of the WZ and ZB QWs are shown in Fig.~\ref{fig:apt}(b). Profiles of the In mole fraction taken along $\langle110\rangle$ directions [Fig.~\ref{fig:apt}(c)] confirm that the In fraction is lower in the WZ QW. However, the FWHM of the QW, extracted by a fit with a Gaussian, is the same ($4.7\pm0.5$ nm) within the experimental error. The variation in QW composition along the length of the NW is plotted in Fig.~\ref{fig:apt}(e), spatially aligned with the CL map in Fig.~\ref{fig:apt}(d). The average In mole fraction in the ZB QW is 15.3\% with a symmetric variance of 0.9\%, which is comparable to the systematic error in the APT measurement of about 1\%. \cite{Rigutti_2016, Riley_2014} In contrast, the WZ segment exhibits continuous gradients in composition from the interface regions to the center of the segment, and these correlate well with the shifts in emission energy seen in Figs.~\ref{fig:intro}(d) and \ref{fig:xrd1}(d), as well as the inhomogeneous broadening discussed further below. 
The minimum of the In mole fraction is $10.7\pm0.1$\%, where the reported uncertainty is determined by counting statistics and is therefore a lower bound. Assuming that any systematic measurement errors are the same for the WZ and ZB segments, we find a maximum difference in In mole fraction of $4.6\pm0.2$\%.

An influence of the underlying polytype on the shell composition and thickness in NWs has previously been reported for different III-V material combinations such as GaAs/InAs,\cite{Rieger_2014} GaAs/(Al,Ga)As,\cite{Jeon_2015} InAs/InP,\cite{Ghalamestani_2012} and GaAs/ Ga(As,Sb).\cite{Ghalamestani_2013} It is evident that (In,Ga)As/GaAs also exhibits this behavior, as the presence of the WZ segment results in a lower In content as compared to the ZB region.

\begin{figure}[t!]

\includegraphics*[width=8.5cm]{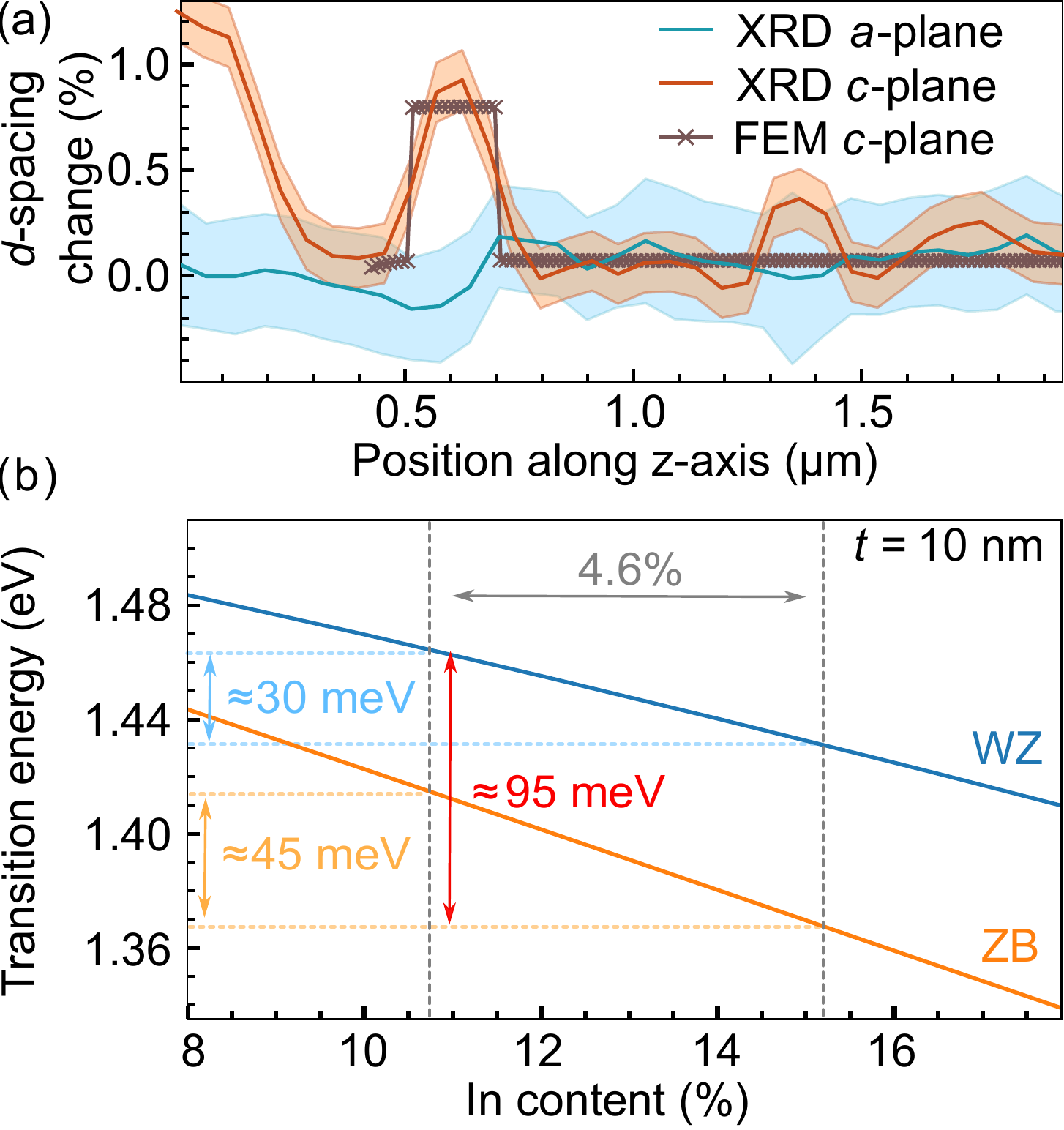}
\caption{\label{fig:sim}(a) Line profiles of the relative d-spacing change (with respect to bulk ZB GaAs) for the \emph{a}- and \emph{c}-plane reflections as obtained by nanoXRD on sample A (NW A3) compared with FEM calculations for the \emph{c}-plane. The experimental error is shown as semi-transparent bands. (b) Transition energies obtained from two-dimensional in-plane $\mathbf{k}\cdot\mathbf{p}$ calculations across (In,Ga)As core-shell heterostructures for varying In content at constant QW thickness $t$. The dashed lines highlight the change in transition energy expected for a change in In content of 4.6\%, amounting to 30--45~meV when ignoring and 95~meV including polytype effects, respectively.}
\end{figure}

\textbf{Simulations of the shift in emission energy.} In order to calculate the difference in emission energy, we must also consider the possible influence of QW strain. Therefore, nanoXRD measurements of the relative $d$-spacing along the NW axis \cite{Holt_2014,Stankevic_2015} were performed on NWs from sample A [example in Fig.~\ref{fig:sim}(a)]. Within the accuracy of the experiment, the $a$-plane spacing [$(2\bar{1}\bar{1}0)/(20\bar{2})$] is unchanged along the whole NW. In contrast, the $c$-plane spacing [$(0002)/(111)$] is increased by $0.9\pm0.2$\% in the WZ segment. Finite element calculations (FEM) based on literature values for the lattice spacing in binary WZ and ZB GaAs NWs\cite{Biermanns_2012} predict a change in $c$-plane spacing of 0.8\%. Hence, we can conclude that the change in c-plane spacing arises primarily from the change in crystal structure, and any residual strain (if present) is below the accuracy limit of our measurement. Further, FEM suggests that changes in the thickness or composition of the (In,Ga)As QW lead to minimal changes of the strain state of the GaAs core (see \textcolor{magenta}{Supporting Information}). Therefore, though the (In,Ga)As QWs are compressively strained from the surrounding GaAs, we conclude that the variation in residual strain between the QWs grown on WZ versus ZB can be neglected.

Now, understanding the composition, morphology, and strain in the QW, as well as the crystal structure of the NW, we can simulate the electronic characteristics of the QWs in the WZ and ZB phase via $\mathbf{k}\cdot \mathbf{p}$ calculations. First, we have independently modeled the electronic properties of the ZB and WZ segments based on their equilibrium lattice constants (see \textcolor{magenta}{Supporting Information} for a careful evaluation of the employed material parameters). Second, for both the WZ and ZB polytype, the composition of the In$_x$Ga$_{1-x}$As shell QW has been systematically varied and the resulting transition energy, defined as energy difference between electron and hole ground state $E(\Psi_\mathrm{el}^0) - E(\Psi_\mathrm{ho}^0)$, is shown in Fig.~\ref{fig:sim}(b). Because excitonic effects and the role of the excitation density are not taken into account, the absolute transition energies are not expected to be reproduced, but the relative changes should be fairly accurate.

Experimentally, we observe a maximum change in In content between the QW on the WZ and ZB segments of $4.6\pm0.2$\%, accompanied by a $75\pm15$~meV shift in the transition energy. According to the calculations, a shift of 30--45~meV could be produced by a 4.6\% change in QW composition, depending on whether we take the band parameters for WZ or ZB [marked by dashed lines in Fig.~\ref{fig:sim}(b)]. However, the $\mathbf{k}\cdot \mathbf{p}$ calculations also predict that an additional 50--65~meV shift results from the difference in band structure between the ZB and WZ polytypes. This is a significant finding, as an interpolation of the experimentally-observed band gaps in binary alloys would predict a shift of only 10~meV in unstrained In$_{0.15}$Ga$_{0.85}$As.\cite{Senichev_2018,Vainorius_2018,Rota_2016} However, for an In content of 15\%, the ternary QWs experience a compressive strain of about 1.1\% from the lattice mismatch with the GaAs core and outer shell. The opposing impact of this strain on the valence band energies in WZ and ZB (In,Ga)As accounts for the remaining shift in energy. Through the crystal field splitting, the heavy hole valence band dominates in WZ (in contrast to the light hole band in ZB) and is shifted to lower energies; a similar effect has been observed when mechanically introducing strain in binary ZB and WZ GaAs NWs.\cite{Signorello_2013,Signorello_2014} In addition, for our QW structures, the different nature of the valence band in WZ also enhances the confinement of the hole states, introducing an additional increase of the transition energy (see \textcolor{magenta}{Supporting Information} for further details).

The combination of the strain effect on the band structure and the change in In content add up to a predicted blueshift of about $95$~meV. The slight overestimation compared with the experiment can be attributed to uncertainties concerning the employed WZ band parameters.

\section{Summary and Conclusions}

In summary, the concurrent growth of (In,Ga)As QWs on adjacent ZB and WZ templates in the form of GaAs NW cores, together with a suite of spatially-correlated single NW measurements, allowed us to establish the influence of the crystal polytype on the growth and emission properties of these ternary QWs. For the WZ QW, we find a reduced In mole fraction and an enhanced influence of strain on the band structure through a shift in the valence band. Both of these factors contribute to the observed blueshift in emission energy. The influence of carrier localization at the stacking defects in the ZB/WZ interface boundary region, as well as spatial variations in In mole fraction explain the inhomogeneous broadening of the WZ QW emission peak. In consequence, the peak centroid shift of  about 40 meV underestimates the difference between the QW transition energies in WZ and ZB, which we estimate to amount to $75\pm15$~meV. Note that this correlative approach can be applied to a variety of open questions concerning the properties of semiconductor nanostructures.

Previously, the emission behavior between WZ and ZB binary GaAs has been studied in detail. However, in NW optoelectronic applications, core-shell heterostructures with ternary QWs are much more relevant. Our correlative analysis deconvolves the effects of composition and strain in ternary (In,Ga)As QWs, showing that the difference in emission energy between the WZ and ZB QW regions is much larger than for binary GaAs as a consequence of the compressive strain on the QW. We thereby establish key design principles for the application of crystal properties in engineering III-As emitters. 

\section*{Methods}

\textbf{Growth.} 
The investigated samples were grown by molecular beam epitaxy (MBE) on p-type B-doped Si $(111)$ wafers covered with the native oxide. The MBE system is equipped with one In and two Ga effusion cells, as well as Be and Si effusion cells for dopants and two valved cracker sources for supply of As$_2$. An optical pyrometer was used to measure the substrate temperature. Fluxes were calibrated and expressed in terms of an equivalent growth rate on the GaAs $(001)$ surface in monolayers per second (ML/s).

Before growth, the Si substrates were annealed in the growth chamber for 10~min.\ at 680\,\celsius. Then, the substrate temperature was adjusted to 645\,\celsius for the growth of the GaAs NW core by the self-assisted vapor-liquid-solid (VLS) method. The core growth\cite{Kupers_2017} was initiated by the deposition of Ga for 30~s at a flux of 1.3~ML/s. After a 60~s ripening step without flux, the Ga and As$_2$ fluxes were supplied simultaneously at a Ga flux of 0.3~ML/s and a V/III flux ratio of 2.2--2.8.\cite{Bastiman_2016} The growth time was 30~min., after which the Ga shutter was closed and the VLS Ga droplets on top of the NWs were consumed by crystallization to GaAs under an As$_2$ flux of 4~ML/s. Subsequently, the substrate temperature was reduced to 420\,\celsius for the lateral shell growth under a V/III flux ratio of 20, conditions that limit adatom diffusion on the NW sidewalls.\cite{Dimakis_2014} Two different samples, A and B, with the following multi-shell structure were grown:

\emph{Sample A:} A full light emitting diode (LED) structure consisting of a radial p-i-n structure. The GaAs NW core is doped p-type using Be and has a diameter of about 100~nm. The multi-shell structure consists of (from the core to the outer shell) a 10~nm thick undoped GaAs shell, a 10~nm thick In$_{0.15}$Ga$_{0.85}$As QW shell, another 10~nm thick undoped GaAs shell and a 50~nm thick n-type GaAs outer shell doped with Si.

\emph{Sample B:} To reduce the NW diameter for the APT measurements, sample~B consists of a 85~nm thick undoped GaAs NW core with a multi-shell structure consisting of a 10~nm thick In$_{0.15}$Ga$_{0.85}$As QW shell and a 20~nm thick undoped GaAs outer shell.

It was verified that the difference in doping and diameter between samples B and A did not influence the crystal structure.

\textbf{Electron backscatter diffraction.}
EBSD in a SEM allows to determine the crystal polytype of extended segments in GaAs NWs.\cite{Lin_2017} To this end, a Zeiss Ultra~55 field-emission SEM equipped with an EDAX/TSL EBSD system was operated at an acceleration voltage of 20~kV and a beam current of 1.7~nA. NWs were dispersed on a Si substrate covered with Au-markers to facilitate subsequent CL measurements on the same NW. The sample was tilted at 70\grad and the axis of the probed NW was aligned along the vertical direction. EBSD maps were acquired with a step size of 20~nm, though the interaction volume of the backscattered electrons and thus the actual spatial resolution is estimated at about 50~nm. The recorded Kikuchi patterns were automatically indexed using the provided routines of the manufacturer. Note that each of the correlations presented in the manuscript was verified on several NWs to ensure that the conclusions we draw can be generalized.

\textbf{Cathodoluminescence spectroscopy.}
CL spectroscopy was carried out in the same SEM, which is also equipped with a Gatan monoCL4 system, at an acceleration voltage of 5~kV and beam currents of about 500~pA. A diffraction grating with 600~lines/mm blazed at 800~nm was used in conjunction with a slit width of 0.5~mm (hyperspectral maps) and 2~mm (monochromatic maps), which corresponds to spectral resolutions of about 4 and 16~meV, respectively. The luminescence was detected by a photomultiplier tube (PMT) for monochromatic maps and by a charge-coupled device (CCD) for hyperspectral maps. Hyperspectral line-scans along the axis of single NWs were collected by recording the luminescence spectra using the CCD at each dwell point of the electron beam. Note that the PMT is close to the limit of its detection range for the probed QW emission and its sensitivity is not corrected for, whereby shifts in the emission energy from NW to NW and between the WZ and ZB segments can have a significant impact on the recorded intensity in Fig.~\ref{fig:intro}(a). The samples were cooled to 10~K using liquid He with a dedicated stage. The correlation to EBSD and nanoXRD was carried out on dispersed NWs from sample A. Prior to the APT measurements, as-grown NWs close to the cleaving edge of sample B were measured with the sample mounted at an angle of 90\grad. The positions were documented by overview micrographs to facilitate harvesting of these very NWs. CL data analysis was performed using routines based on the python library hyperspy.\cite{Pena_2018}

\textbf{Nanoprobe x-ray diffraction.}
Nano-diffraction measurements were performed at the Hard X-ray Nanoprobe beamline 26-ID-C of the Advanced Photon Source at Argonne National Laboratory. Simultaneous x-ray fluorescence mapping of the Ga K-edge (10.367 keV) was used to locate the NWs, which were randomly dispersed on a 10~$\upmu$m thick Si substrate transparent to hard x-rays. A nanofocused beam of $\approx25$~nm diameter was scanned across the NWs, and diffraction patterns for a specific Bragg reflection were measured on a 2D CCD detector placed 0.8--0.9~m away from the sample. A total of eight NWs were measured at 2--3 different Bragg conditions each. NWs oriented in two different directions on the substrate were measured, which gave access to different Bragg reflections. The first set of NWs, which includes NW A2 from Fig.~\ref{fig:xrd1}, were investigated at the \emph{a}-plane condition of $(2\bar{1}\bar{1}0)$ and the \emph{m}-plane $(10\bar{1}0)$ reflection (forbidden for ZB). The second set of NWs, including NW A3 from Fig.~\ref{fig:sim}(a), was investigated at the same $(2\bar{1}\bar{1}0)$ reflection, but also at the \emph{c}-plane condition of $(0002)$ and the $(10\bar{1}1)$ reflection, which has components of both \emph{c} and \emph{m}-planes (forbidden for ZB). Whereas the first set of NWs did not allow access to the \emph{c}-plane reflection, they gave a better view of the stacking order using the \emph{m}-plane reflection. 
 
\textbf{Atom probe tomography.}
Samples for APT were prepared by a multi-step transfer method in a SEM (see \textcolor{magenta}{Supporting Information} for images).\cite{Koelling_2017} First, a sacrificial NW was Pt welded to the end of a tungsten micromanipulator tip. The sacrificial NW was then welded to the base of the NW of interest that was standing vertically (as grown) on the edge of a Si substrate. The NW was then pulled off the substrate, the SEM was vented, and the manipulator tip was rotated by hand to tilt the vertically standing NW by approximately 30\grad. The NW was then Pt welded to the top of a tungsten tip standing vertically in the SEM and the sacrificial NW was broken off. The NWs were then cleaned in oxygen plasma for 3~min. The NW diameter of 145~nm (short axis) is still significantly larger than common for APT tips. In this case, mounting to tungsten tips was essential for the analysis of the NW without fracture, which was not possible using a traditional Si micropost array. APT measurements were performed in a LEAP 5000XS system utilizing a 355~nm laser, at a 30~K stage temperature, 250~kHz laser pulse rate, and a 2\% detection rate. A pulse energy of 10~pJ was used to evaporate through the overgrowth layers, and then was lowered to 1~pJ, with the specimen reaching a maximum of 2.5~kV. The APT measurements are consistent for a total of three analyzed NWs.

\textbf{Simulations.} 
Calculations based on $\mathbf{k}\cdot\mathbf{p}$ perturbation theory are well adapted to assess the properties of multiple conduction and valence bands around the center of the Brillouin zone with a reasonable computational effort.\cite{LewYanVoon_2009} Elastic and electronic properties were computed using the respective modules of the Sphinx software library.\cite{Marquardt_2014} In this framework, we employ an eight-band $\mathbf{k}\cdot\mathbf{p}$ model for WZ semiconductor materials.\cite{Chuang_1996} As the segments are long enough that we can ignore the influence of the interfaces on their elastic properties, the calculations have been restricted to 2D in-plane cuts through the NW [c.f.\ sketch in the inset to Fig.~\ref{fig:intro}(a)]. The calculated structure corresponds to that of sample B. The parameters of the ZB phase were transformed to the ones of the WZ Hamiltonian using the relations provided in Ref.~\citenum{Climente_2016}. All relevant parameters employed are summarized in the \textcolor{magenta}{Supporting Information} (using the WZ notations) together with a discussion on their reliability. Elastic strain was computed using a linear elasticity model and enters the eight-band $\mathbf{k}\cdot\mathbf{p}$ model in a similar manner as in Ref.~\citenum{Marquardt_2017}. Excitonic effects were neglected in this work and thus represent a systematic error (of about 5--10~meV\cite{Fu_1991}) to be considered when comparing simulation results to transition energies observed in experiment.

\textbf{Transmission Electron Microscopy.}
NWs from sample B were investigated by high-resolution TEM 
in a JEOL 2100F field emission microscope operated at 200~kV and equipped with a Gatan Ultra Scan 4000 CCD camera for image recording. For this investigation, the NWs were mechanically dispersed on a Lacey carbon film supported by a 300 mesh copper grid. Each investigated NW was aligned along the $\langle 11\bar{2}0\rangle$ WZ zone axis (corresponding to the $\langle 1\bar{1}0\rangle$ ZB zone axis) to obtain high resolution micrographs and a clear contrast from SFs.

\section{Associated Content}

\textbf{Supporting Information.} Panchromatic CL map; illustration of nanoXRD scattering geometries; FEM calculations of NW strain state; additional CL line-scan for sample B; NW harvesting for APT measurements; video animation of the APT data; extended discussion of the $\mathbf{k}\cdot\mathbf{p}$ calculations.

\textbf{Author Contributions.} 
JL and MOH contributed equally to this work in design, experimentation, analysis, and writing of the manuscript, with contributions from LJL and LG to overall conceptualization. UJ started the study and contributed to CL and EBSD measurements led by JL, while JH grew the samples with input from JL and LG. MOH led experiments and analysis of APT and nanoXRD measurements. LJL, AD, AAH, UP, SOH, and MVH contributed to the design and analysis of the nanoXRD investigation, and AD, AAH, SOH, MVH, IC-A and CH contributed to the nanoXRD measurements. AAH carried out FEM simulations. LJL, CH, and MOH designed APT experiments. GG did the TEM measurements. OM developed the $\mathbf{k}\cdot\mathbf{p}$ model in discussion with JL, JH and LG. All authors contributed to understanding of correlative measurements and with manuscript editing. 

\textbf{Acknowledgements.} 
The authors would like to thank Hanno K\"upers and Oliver Brandt for discussions and Michael Hanke for a critical reading of the manuscript. The authors are grateful to M. H\"oricke and C. Stemmler for MBE maintenance. The authors are also grateful for assistance from M.J. Moody and Z. Sun in performing nanoXRD measurements. OM acknowledges support from the DFG through SFB 787. LJL and MOH acknowledge support of NSF DMR-392 1611341. MOH acknowledges support of the NSF GRFP. X-ray nanodiffraction experiments and data reduction was supported by the U.S. Department of Energy (DOE), Office of Basic Energy Sciences (BES), Materials Science and Engineering Division. The nanodiffraction measurements were performed at the Hard X-ray Nanoprobe beamline 26-ID-C operated by the Center for Nanoscale Materials and Advanced Photon Source at Argonne National Laboratory. Use of the Center for Nanoscale Materials and the Advanced Photon Source was supported by the U.S. Department of Energy, Office of Science, Office of Basic Energy Sciences, under Contract No. DE-AC02-06CH11357. This work made use of the EPIC facility of the NUANCE Center at Northwestern University, which has received support from the Soft and Hybrid Nanotechnology Experimental (SHyNE) Resource (NSF ECCS-1542205); the MRSEC program (NSF DMR-1720139) at the Materials Research Center; the International Institute for Nanotechnology (IIN); the Keck Foundation; and the State of Illinois, through the IIN.

\bibliography{GaAs-WZ-ZB}

\end{document}